# Urban association rules: uncovering linked trips for shopping behavior

Yuji Yoshimura[a,b], Stanislav Sobolevsky[a,c], Juan N Bautista Hobin[a], Carlo Ratti[a], Josep Blat[b]

[a] SENSEable City Laboratory, Massachusetts Institute of Technology, 77 Massachusetts Avenue, Cambridge, MA 02139, USA;

[b] Department of Information and Communication Technologies, Universitat Pompeu Fabra, Roc Boronat, 138, Tanger Building 08018 Barcelona, Spain;

[c] Center for Urban Science and Progress, New York University, 1 MetroTech Center, 19th Floor, Brooklyn, NY 11201, USA;

**Abstract.** In this article, we introduce the method of urban association rules and its uses for extracting frequently appearing combinations of stores that are visited together to characterize shoppers´ behaviors. The Apriori algorithm is used to extract the association rules (i.e., if -> result) from customer transaction datasets in a market-basket analysis. An application to our large-scale and anonymized bank card transaction dataset enables us to output linked trips for shopping all over the city: the method enables us to predict the other shops most likely to be visited by a customer given a particular shop that was already visited as an input. In addition, our methodology can consider all transaction activities conducted by customers for a whole city in addition to the location of stores dispersed in the city. This approach enables us to uncover not only simple linked trips such as transition movements between stores but also the edge weight for each linked trip in the specific district. Thus, the proposed methodology can complement conventional research methods. Enhancing understanding of people's shopping behaviors could be useful for city authorities and urban practitioners for effective urban management. The results also help individual retailers to rearrange their services by accommodating the needs of their customers' habits to enhance their shopping experience.

## 1. Introduction

In this paper, we explore the applicability of association rules (Agrawal et al., 1993) for extraction of combinations of visited stores for the analysis of linked trips for shopping behavior. The Apriori algorithm (Agrawal & Srikant, 1994), which is widely used and a rather simple yet robust method for market-basket analysis, is applied to our anonymized large-scale transaction dataset. This algorithm was originally designed to extract combinations of other items most likely to be purchased by a customer given a particular item that is already in his or her basket as an input. Instead of the purchased items in a single store, we try to extract the links between the stores in which people make transactions before or after visiting some focal shops, considering all transactions conducted in stores dispersed throughout the city. Thus, we can uncover the edge weight for each trip as the transition probability between stores by comparing it with the general pattern of all other shoppers' behaviors in the given district.

An anonymized bankcard transaction dataset provides us with longer-term evidence that people perform transactions among stores as a digital footprint or "data exhaust" (Mayer-Schönberger & Cukier, 2013, p113). Unobtrusive observations (Webb et al., 1966) can be used to reconstruct customers' sequential movement between the stores in which they make purchases. This point makes our analysis different from previous studies, which have only attempted to "infer" people's activities during their trips and use them for linked trip analysis. For example, the datasets obtained from individual-based global positioning system (GPS)-enabled smartphones are helpful for imputation processes, together with land-use data, to infer people's activities (Shen & Stopher, 2013). Media access control (MAC) address detection techniques such as Wi-Fi and Bluetooth can also be useful to detect the patterns of users' activities between stable places and infer the type of regular activity depending on place, the day of the time and season (Yoshimura et al., 2014; Yoshimura et al., 2016).

Although these analytical methodologies can help alleviate the shortcomings of traditional travel surveys (Rasouli & Timmermans, 2014; Shoval & Issacson, 2006; Stopher & Shen, 2011; Bricka et al., 2012) and greatly help in behavioral analysis, none of them can generate information regarding actual people's expenditures and their successive movements between stores for a longer term throughout a city.

This paper aims to complement to the existing methods for analyzing linked shopping trips. The travel survey methods have provided the fundamental data for capturing personal travel characteristics, which have a proven value (see Weiner, 1997 for a review). However, they are not adequate for monitoring the long-term characteristics of non-work/school trip behaviors. Transaction datasets can, though, provide us with continuous and long-term travel information. Conversely, the latter's trip information is quite fragmentary and does not contain sufficient information about the socio-demographic characteristics of travelers, their trip purposes, or the transportation mode, which the former can typically provide. Alternatively, the combination of GPS technology and interviews or questionnaires can uncover the visited stores and customers' paths, attributions, visit motivations, and expenditures, but this methodology can be applied only in spatially and temporally limited terms (e.g., only for a shopping street over a time span of a few days) due to the larger burden for respondents. The mentally demanding and time-consuming method results in small-scale samples. Thus, our proposed methodology can shed light on another aspects of shopping behavior that the conventional analysis cannot achieve.

We begin with an introduction of association rules and their potential contribution to linked trip analysis. Section 2 gives a brief discussion of association rules, and Section 3 describes the study area, the data-collection technique, and the analytical framework. In Section 4, we will apply urban association rules to identify the linked trips for the shopping behaviors. Finally, conclusions are drawn in Section 5.

## 1.1 Mobility and linked trip study: collecting data from digital footprints

Shopping can be considered one of the most important activities for generating linked trips. Some of the characteristics of shopping behaviors are follows: They are highly varied and change in both the spatial and time components, which makes shopping differ from daily activities such as work/school-related ones (Wang & Miller, 2014). One of the aspects of shopping, "the multipurpose and the purchasing of different items on a single trip" (Arentze et al., 2005), indicates that traditional household surveys such as person trip surveys or usual single-day trip surveys may have difficulty capturing shopping behaviors for behavioral analysis because the aforementioned methodology is based on trip-based data collection.

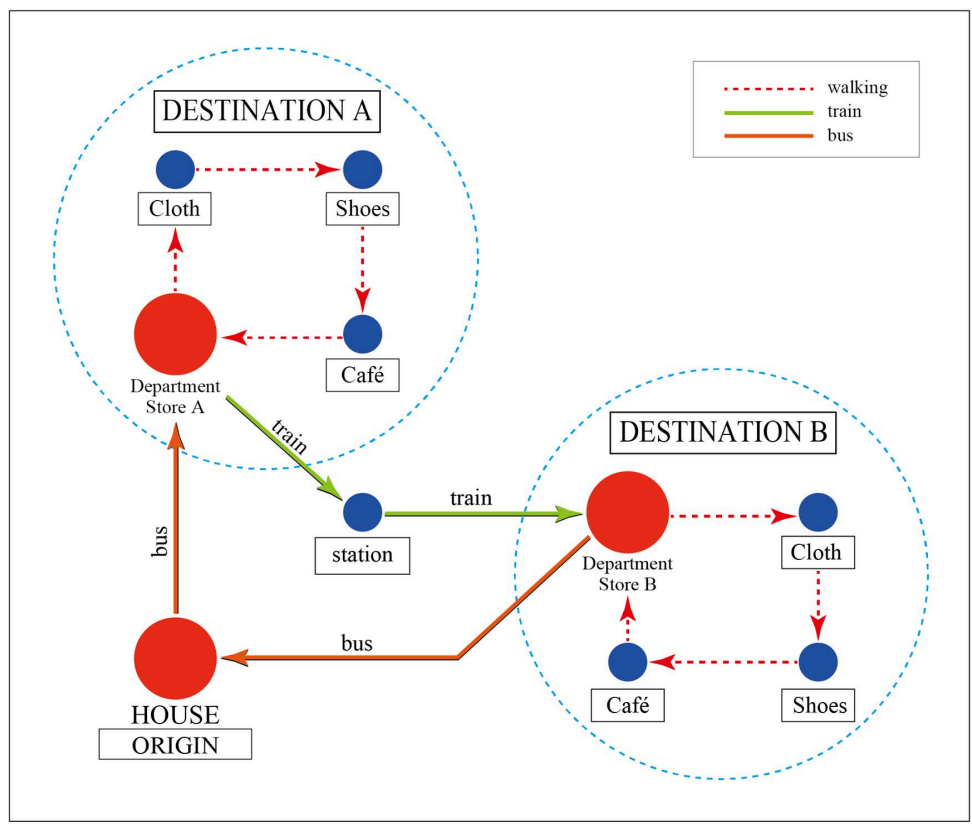

**Figure 1**. An example of linked trips during a day.

Figure 1 presents the example of linked trips during a day. All arrows represent the real trip of this person, but it is known that people tend not to report all trips, especially short trips around a destination (dotted red lines). Thus, his/her trip is likely to be represented only by the orange and green arrows. The problems regarding the conventional questionnaire-type travel surveys were discussed in Axhausen (1998). To overcome them, the activity-based approach is proposed to better understand people's travel behaviors and the decision structure underlying them, considering time use patterns (for an overview, see Ettema & Timmermans, 1997).

Timmermans and his co-workers performed important research regarding linked shopping trips in the framework of modeling and analyzing the dynamics of pedestrian behavior (Borgers & Timmermans,1986a; Borgers & Timmermans,

1986b; Kurose et al., 2001; Borgers & Timmermans, 2005; Zhu & Timmermans, 2008; Timmermans, 2009; Dijkstra et al., 2009; Kemperman et al., 2009; Dijkstra et al., 2014).

Their methodology largely relies on on-site interviews and questionnaires to ask people at the entrances and exits of the shopping district about the following: the locations of the shops that they visited, their route choices, their expenditures, the starting and ending times of their shopping trip, and their mode of transport to the city center. Information about individual socio-economic and demographic variables and psychological factors is also collected. They sometimes complement these traditional qualitative methods with state-of-the-art technology (Axhausen et al., 2002). Comparative studies of GPS and travel survey data are discussed in Bricka & Bhat (2006), and a summary of the modeling of shopping destination choices is presented in Huang & Levinson (2015).

Those combinations enable us to collect human behaviors on a finer granular scale in space and time and greatly improve the quality of datasets for travel analysis (Rasouli & Timmerman, 2014). However, they are also not error-free. We identified the following points as the shortcomings of the previous studies, particularly for linked trips for shopping behaviors.

First, individual-oriented research methods, such as *people-centric sensing* (Miluzzo et al., 2008) [in contrast with methods conducted manually or that employ emerging technologies] can merely collect information about sample shopping activities in the specific area where data collection is performed. This raises the significant question: whether the observed behavior is unique to the district in which the sample was collected, - e.g., due to the characteristics of the distribution of stores, their types and their number (retail agglomeration) – or is independent of those environmental factors. The conventional research method is not suitable to address this question due to the shortcomings of datasets obtained using this data collection methodology (e.g., interviews or observations).

Second, the dataset used for people's linked trips is both temporally and spatially limited, as we described previously. The sample data collection is typically conducted to intercept pedestrians at the entry/exit points of shopping areas over one day or a few days, resulting in a small number of samples. The results are ineffective for analyzing repetitive choice behavior and assessing the temporal effects of various policies.

Finally, there is a lack of robust tools to analyze the large-scale dataset of people-linked trips. As we described above, the incorporation of emerging technologies in the data collection process enables us to capture human behavior on finer granular scales in space and time and to increase the scale of the quantity of datasets. However, the increasing number of relevant datasets is not only expected to provide new sources for human behavioral analysis but also to cause the problems in the analytical process. We are familiar with analyzing small-scale samples but not yet prepared for large-scale datasets.

We try to complement the shortcomings of the above-mentioned methodology for the analysis of linked shopping trips. We propose to use transaction datasets consisting of more than 100 million unique users and present an adequate methodology for analyzing such datasets. This enables us to analyze the temporally ordered origins and destinations of customers' spending behaviors all over the city rather than in limited areas, such as shopping districts. This unique dataset and its analytical methodology also make it possible to compute the number of stores classified into categories in each area and the number of transactions made in each category in each area. Thus, we can uncover not only the simple transition movements between the stores at which customers made transactions but also the edge weight for each trip (link) by computing the ratio with all other relevant transactions in each district.

Our proposed methodology can greatly enhance our knowledge about people's shopping behaviors in the city. The results can provide a distribution map of people's shopping activities and the composition of stores. This is useful for urban managers and city authorities to identify the unevenness of the spatial distributions of people's shopping activities. In addition, our approach can work as a fast, inexpensive and robust filtering system to capture the significant variables within the large-scale datasets. This greatly reduces costs before starting to explore the causality in a whole dataset. Thus, our proposed methodology is more useful for analysis based on a large-scale dataset rather than analysis based on small-scale samples.

## 2. Methodology: from *association rules* to *urban association rules*

The analytical framework for this paper is based on association analysis and classification of the obtained results. We analyze the correlations among stores where transactions were made as discrete variables, resulting links between them. This is different from uncovering the causality by relying on randomly chosen small-scale samples, which can be most likely be best obtained from qualitative travel survey data or diary surveys that can elucidate the behavioral process in detail. Such studies also attempt to capture activity scheduling and rescheduling processes (Arentze & Timmermans, 2000; Timmermans, 2001; Miller & Roorda, 2003; Arentze & Timmermans, 2005; Nijland et al., 2009) because the trip could be the consequence of decisions at an earlier stage of behaviors. Hence, activity-based approaches can be used to overcome some of the typical shortcomings and limitations of a personal trip survey (Timmermans, 2005).

Conversely, the present work explores the correlations between the stores that customers made transactions by increasing the quantity of data because "when we increase the scale of the data that we work with, we can do new things that weren't possible when we just worked with smaller amounts" (Mayer-Schönberger & Cukier, 2013, p10).

Association analysis was proposed by Agrawal et al., (1993) to extract interesting relationships between items purchased by a customer in a market basket by using historical transaction datasets collected in stores (see Tan et al, 2005, pp327-414 for a review). The objective of this analysis is to extract combinations of items that frequently appear in the transaction datasets and seek hidden patterns embedded in the large-scale datasets.

**Table 2.** An example of market-basket transactions made based on Table 6.1. from Tan et al. (2005)

| TID | Items |
|---|---|
| 1 | {Bread, Fruit, Jam} |
| 2 | {Diapers, Milk, Beer, Ham, Salad} |
| 3 | {Milk, Diapers, Beer, Fish, Salad} |
| 4 | {Fruit, Diapers, Bread, Milk, Jam, Salad, Beer} |
| 5 | {Bread, Milk, Diapers, Salad, Jam} |

Table 2 presents an example of a set of items purchased by customers in a grocery store. Each row corresponds to a transaction by a given customer, which contains a unique identifier, i.e., a transaction ID (TID), and a set of items. For example, the following rule can be extracted from the table of transactions:

$$\{Diapers\} \rightarrow \{Beer\} \qquad (1)$$

This is because many customers who purchase diapers also buy beer, thus indicating that there exists a strong relationship between the sales of these two items. The support count, $\sigma(X)$ for an itemset X indicates the number of transactions that contain a particular itemset. For instance, in Table 2, the support count for {Diapers, Salad, Beer} is 3 because there are only three transactions that contain all three items.

Association rules are rules that surpass the minimum *support* and minimum *confidence* thresholds defined by a user. The strength of an association rule can be measured by the *support, confidence* and *lift*.

$$\text{Support, } s(X \rightarrow Y) = \sigma(X \cup Y)/N \qquad (2)$$

$$\text{Confidence, } c(X \rightarrow Y) = \sigma(X \cup Y)/\sigma(X) \qquad (3)$$

$$\text{Lift, } (X \rightarrow Y) = \sigma(X \cup Y)/(\sigma(X)\sigma(Y)) \qquad (4)$$

*Support* determines the frequency of the transaction that satisfies *X* and *Y* out of all of the transactions (N). If a rule has lower support, it might be happening simply by chance. Conversely, *confidence* is the conditional probability that Y (consequent) occurs given that X (antecedent) has happened. A higher value of *confidence* indicates a higher reliability of the inference made by a rule (i.e., $X \rightarrow Y$). That is, the higher the confidence, the more likely it is for Y to be present in transactions that contain X. *Lift* is the fraction of supp (X U Y)/(supp (X) supp (Y)). Greater *lift* values indicate stronger associations.

The concept and methodology of *urban association rules* is based on *association rules*. The significant difference is that whereas the former addresses the items purchased in an individual store, the latter focuses on the stores in which a customer completed a purchase and his/her successive purchasing activities among those shops dispersed in the urban settings. In *urban association rules*, an individual shop is analogous to an item in association rules.

For this paper, SUPPORT is the ratio of the number of transactions in a category of the shop to the total number of transactions in the district. CONFIDENCE indicates the ratio of the number of transactions conducted in a particular category of the shop over the total number of transactions in all categories of the shop. LIFT is the fraction of B/A, where B= the number of transactions made in a particular category of the shop over the total number of transactions made in all categories in the same shop and A = the number of transactions made in the category in the district over the total number of transactions conducted in the district. We can interpret the obtained scores as follows: If the score is 1.0, the customers' behavior in the shop and the general public's behavior in the district are similar. If the score is extremely higher or lower (i.e., >1.5, <0.5), the behaviors of those two groups are quite different. Thus, for this paper, LIFT reveals the unique feature of the targeted shops' customers' transaction activities in terms of the particular category, compared with the general consumption behavior of people in the district.

## 3. Study area and dataset

The present study employs the complete set of bank card transactions recorded by Banco Bilbao Vizcaya Argentatia (BBVA) during 2011 from throughout Spain. This study built on the earlier studies about the nature of datasets obtained through analysis of customers' spending behaviors. Sobolevsky et al., (2015a) examined a relationship between the sociodemographic characteristics of customers (e.g., age and gender) and their spending habits and mobility. Sobolevsky et al., (2015b) also researched how Spanish cities attract foreign customers by analyzing transaction activities.

### 3.1 Study area

We analyzed people's linked trips for purchase behaviors in the city of Barcelona, Spain. Barcelona is the capital of the autonomous community of Catalonia, which is located on the Mediterranean coast in northeastern Spain. Approximately 1.6 million people inhabit 100 square kilometers of land with a density of 159 hab/Ha. The city is surrounded by two rivers (the Llobregat and Besòs Rivers), a mountain (Collserolla) and the Mediterranean Sea.

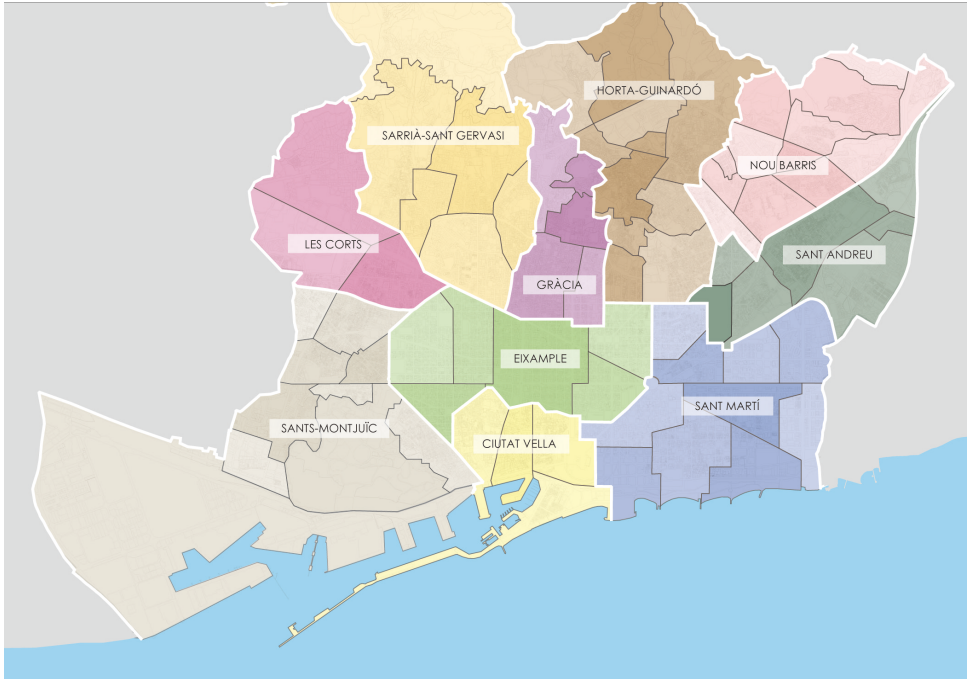

**Figure 2.** A map of the city of Barcelona. There are 10 districts, which contain 73 neighborhoods.

The city of Barcelona is divided into 10 administrative districts, and 73 neighborhoods within those districts, each of which has its own identity (see Figure 2). There are approximately 50,000 locations for economic activity throughout the city.

For the analysis in this paper, we chose the customers who made transactions in three department stores that are located in different neighborhoods of the city (i.e., the shops h, k, and j). We selected the department stores as the focal shops for the analysis rather than small- or medium-scale shops because (1) each one can be considered to be the strongest hub for shopping in each neighborhood, (2) we can expect a larger number of customer transactions due to the stores' higher attractiveness compared with small- and medium-scale retail shops, and (3) the composition of customers obtained from such shops can be more homogeneously distributed than when considering smaller retail shops.

We define the customers, who made transactions in any other shops before making transactions in any of the shops h, k, or j as incoming customers. Similarly, we define customers, who made transactions in any other shops after making transactions in any of the shops h, k, or j as leaving customers.

The association rule to be tested in this paper can be stated as follows:

$$\{\text{somewhere}\} \rightarrow \{\text{the shop h or k or j}\} \qquad (4)$$

$$\{\text{the shop h or k or j}\} \rightarrow \{\text{somewhere}\} \qquad (5)$$

Thus, we extract the consecutive transaction activities of incoming/leaving customers for the anchor shops h, k and j. Also, we define the "people" in the district as all people who made transactions in the district. This includes the customers of the focal shop that is located in the corresponding district.

## 3.2 Sample characteristics

The datasets used in this paper consist of anonymized bank card transaction data from two groups of card users: bank direct customers who are residents of Spain and hold a debit or credit card issued by BBVA and foreign customers who hold credit or debit cards issued by other banks and made transactions through one of the approximately 300,000 BBVA card terminals in Spain. For this paper, we mainly focus on this first group, customers who are residents of Spain, and in particular those who made transactions in the city of Barcelona during 2011. This results in 4.9 million transactions, which are the target of our analysis.

The data contain randomly assigned IDs for each customer connected with certain demographic characteristics, an indication of a residence location, the shop ID where a customer made a transaction, the date of the transaction, and the amount of money spent. The data were anonymized prior to sharing in accordance with all local privacy protection laws and regulations. Thus, all information that would enable us to identify a cardholder was removed.

The dataset was originally classified into 76 categories, which were further aggregated into 6 general categories with 13 sub-categories (see the Appendix). This was performed based on a previous study about the economic activities for the city of Barcelona (see Porta et al., 2012 for details). In addition, all of the general categories and sub-categories are grouped into daily and non-daily use categories by the following definition: The non-daily use category contains shops "which, in themselves, bring people to a specific place because they are anchorages", whereas the daily use category contains "enterprises that grow in response to the presence of primary uses, to serve the people the primary uses draw" (Jacobs, 1961, pp.161-162).

"Secondary activities are therefore local in scale and service in type" (Porta et al., 2011, pp6). Thus, most of the small-scale retails shops and services are classified into the general category A (retail commerce), which is the daily use activity. We further split this general category A into three sub-categories. The first is A1, which is more oriented to the ordinary needs of the neighborhood (e.g., a vegetable shop). The second is A2, which is more oriented to the specialized economic activities than A1 is (e.g., a jewelry shop). The last one is A3, which is related to motor vehicles.

This classification enables us to analyze how customers visit shops in the daily use and non-daily use categories and their consecutive transactions as linked trips. Thus, our methodology permits us to uncover the relationships in customers' purchasing behaviors between both types of shops, thereby revealing the characteristics of the districts and neighborhoods of the city of Barcelona.

## 4 Results

## 4.1 Analysis of supply of activity locations

The city of Barcelona is composed of approximately 50,000 stores. They are unevenly distributed between districts. For instance, while more than 20% of the total number of stores are concentrated in the Eixample district, only 4.7% are located in the Les Corts district. This distribution is independent of the dimension of the district ($r^2=0.0009$) but slightly associated with the population of the district ($r^2=0.53$).

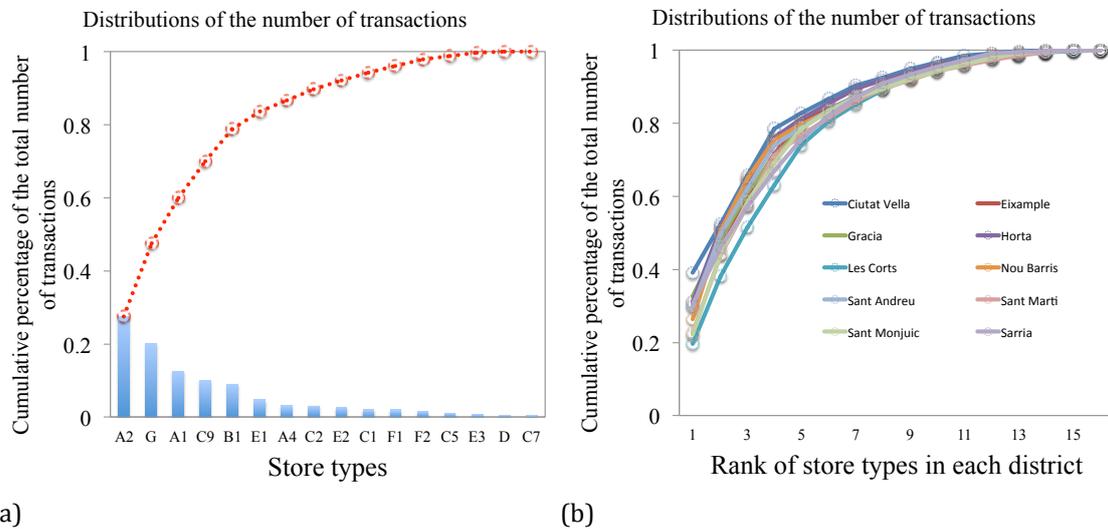

(a) (b)
**Figure 3.** (a) The cumulative distributions of the number of classified stores for the entire city. (b) The cumulative distributions of the number of classified stores for each district.

Figure 3 (a) presents the cumulative distribution of the number of stores classified into the category for an entire city. 27% of all stores are allocated to A2 (comparison goods) and 20% are G (vacant stores); thus, almost half of the total stores are composed of only these two categories. Together with three other categories (A1, C9, and B1), the total number of stores belonging to those categories increases to almost 80%. This indicates that the distribution of the number of stores in each district is largely distorted into a few categories. In addition, the cumulative distribution ordered by the categories that have the larger number of stores is quite similar between districts [see Figure 3 (b)]. The largest and second largest categories are always the same in every district except Ciutat Vella: A2 and B1 are the largest for Ciutat Vella, while A2 and G are the largest for other districts.

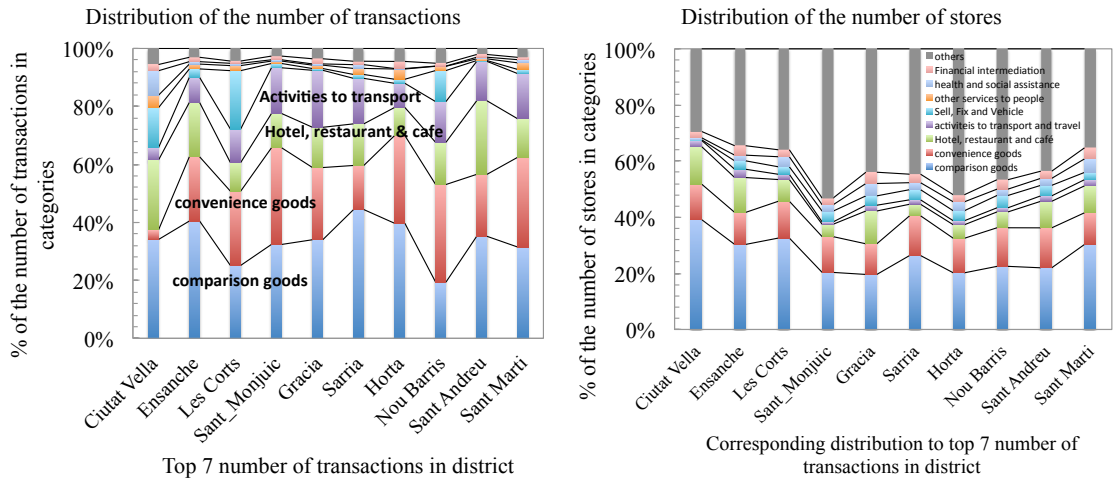

**Figure 4.** (a) Distributions of the number of transactions of the top 7 categories in the districts. (b) Distributions of the number of stores corresponding to the categories in (a).

However, the distribution of the number of stores in each district is quite different from the distribution of the number of transactions made in each district. Figures 4 (a) and (b) show that a larger number of stores does not necessarily result in a larger number of transactions. In order to uncover the supply/demand relationship, we computed the fraction of two factors: the percentage of the total floor space for the store in each district against all districts/the percentage of the number of transactions made in each district against all districts (Figure 5).

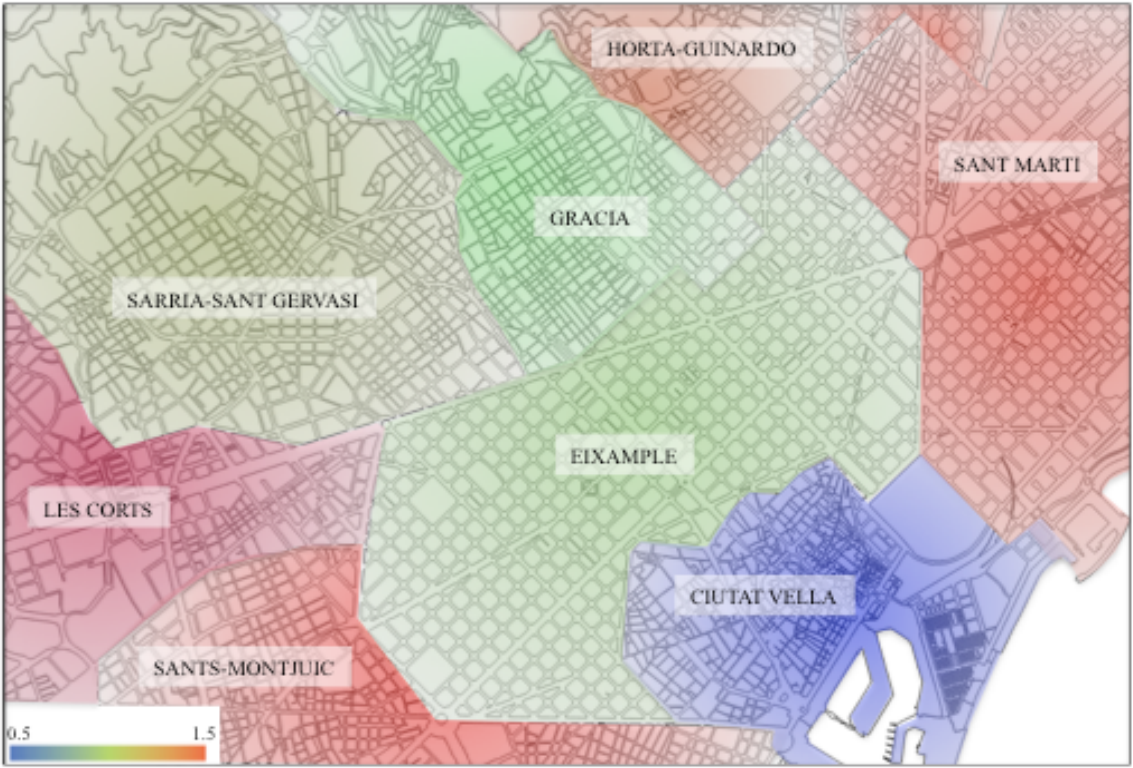

**Figure 5.** Visualization of the ratio of the floor space and the number of transactions made in shops in the city. A higher score is indicated by red and a lower score by blue.

Only the Ciutat Vella district presents a lower value (i.e., 0.55), suggesting that the demand (the number of realized transactions) is superior to the supply (the floor space) in the district. 8 of 10 districts show 0.7<x<1.3, indicating that the volumes of supply and demand are quite well balanced. However, the Sants-Monjuic district has much higher scores (i.e., 1.56), meaning that the number of transactions is much smaller than the floor spaces that exist in each district.

The obtained results are the basis for the following analysis regarding the focal stores' urban association rules. We attempt not only to extract the individual's linked trips to determine the shopping behaviors around each focal store but also to compare each of them with the general patterns of people's shopping behaviors in each district. Thus, we try to uncover the weight of each linked trip conducted by the individual customer around the focal shop in each district.

## 4.2 Linked trips defined by urban association rules

The linked trips for shopping behaviors in the city of Barcelona were evaluated by three indicators of urban association rules: CONFIDENCE, LIFT, and SUPPORT. Each of them captures one of the aspects of the customers' linked trips for the shopping behaviors. Thus, further classification will be conducted by considering all values computed using these three indicators. For the purpose of analyzing all values and their relationships in detail, this paper does not apply the minimum support and minimum confidence thresholds, which are normally defined in association rules (see section 2.2).

Figures 6 (a), (b) and (c) represent graphic examples of CONFIDENCE, LIFT, and SUPPORT, respectively, for the consecutive transactions made by customers around three focal stores (i.e., shop h, shop k, shop j). These figures can be interpreted as follows: the red circle on the left side represents the categories that each focal shop's customers come from, and the blue circle on the right side describes the categories that the customers transit to make transactions after making the transaction in the focal shop. The size of the circle represents the relative magnitude compared with the green circle (=1.0). Considering the general patterns of people's shopping behaviors in each district in which the focal store is located, this representation allows us to depict the features of customers' behaviors with each focal shop.

### *4.2.1 CONFIDENCE and LIFT*

**Figure 6.** (a) (b) (c) Visualization of CONFIDENCE and LIFT. The red color shows the incoming customers, and the blue color presents the leaving customers.

Figure 6 (a), (b), and (C) illustrate differences in the linked trips for shopping behaviors in terms of CONFIDENCE and LIFT.

The transaction patterns of the customers of shops h and k are illustrated in figures 6 [(a), (b)]. The largest number of the customers of each focal shop derives from the category A2 (comparison goods), and moves to the same category. Conversely, we can find a quite different pattern in shop j's customers' activities [Figure 6, (c)]. Whereas the largest number of their customers come from B1 (hotel, B&B, hostel and restaurant, pub and café), the largest number of them move to category A2 (comparison goods). This is a significant difference in terms of the origin-destination of the customers' linked trip for shopping behavior. For the former group, the focal store functions only as a transit location for their continuous activities of movement between the same category, but for the latter, the shop j is the place where customers' transaction activities change in terms of their visited categories before and after visiting this focal store.

However, the larger quantity of customers' linked trips showed by CONFIDENCE does not necessarily indicate the unique characteristics of the customer's transaction activities in each store because it might derive from the general pattern of people's shopping behaviors in the district where the focal shop is located. Thus, we examined LIFT [the right panels of Figure 6 (a), (b), and (c)].

The visualization of LIFT in Figure 6 clearly reveals that the j's customers present extremely higher values of LIFT. This is especially true for the categories of C5 (real estate), A1 (convenience goods) and F1 (recreational). Because the higher score of LIFT indicates a greater difference in transaction activities between the customers and people in the district, the above-mentioned categories of the shop j can be considered unique characteristics of their customers' activities.

Contrary to these facts, the LIFT of the shop h is almost 1.0 in most of the categories [see the right panel of Figure 6 (a)]. This indicates that shop h's customers' transaction activities are quite similar to the people's activities in the district where shop h is located.

All of these analyses would not be possible if we had focused solely on the transitional probability based on consecutive transactions between stores. Such results are significant when we try to explain shopping behaviors and manage the district.

### 4.2.2 Patterns of customers' linked trips

The examination of the scoring of CONFIDENCE and LIFT generated five well-defined groups of customers' behaviors distinguished by the scores of both indicators. The categories that have a higher or lower score were recorded as characterizing the group as a whole. We applied the systematic classification of their behaviors. First, we examine the magnitude of the score of CONFIDENCE and LIFT together. Then, within the identified group, who hold both a higher score of CONFIDENCE and a higher or lower score of LIFT, we study the magnitude of the score of SUPPORT.

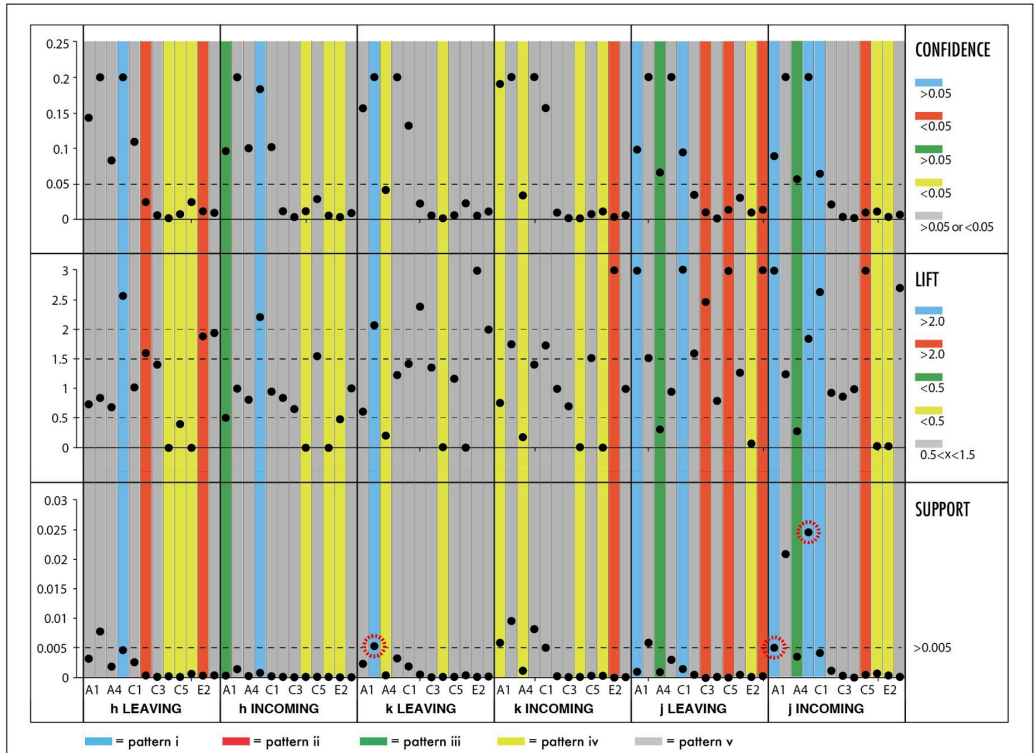

**Figure 7.** Visualization of classification of the customers by three indicators. We classify them into 5 types: (i) an extremely high value of LIFT (>2.0) with a high value of CONFIDENCE (>0.05), (ii) an extremely high value of LIFT (>2.0) with a lower CONFIDENCE (<0.05), (iii) an extremely low value of LIFT (<0.5) with a high value of CONFIDENCE (>0.05), (iv) an extremely low value of LIFT (<0.5) with a low value of CONFIDENCE (<0.05) and (v) a LIFT value of almost 1.0 (0.5<x<1.5) with higher or lower CONFIDENCE (>0.05 or <0.05).

Figure 7 is a graphic of the five groups of customers: the blue color is pattern i, the red color is pattern ii, the green color is pattern iii, the yellow color is pattern iv and the grey color is pattern v (see the detailed classification in the Appendix).

Pattern i indicates that people visiting each district rarely made transactions in the indicated categories, but the customers of the focal shop made consecutive transactions in those categories after or before making transactions in the focal shop. In addition, the number of such customers of each focal shop who made the consecutive transactions in said categories is substantially higher. Conversely, the pattern ii indicates that the number of such customers is not significantly greater. Thus, pattern ii cannot be considered as a unique characteristic of the customers' behaviors in each focal shop.

The pattern iii indicates people who made transactions in each district, and the focal stop's customers behave in quite a different manner. In addition, the number of customers of each focal shop who made consecutive transactions in the indicated categories can be considered as not by chance because of the higher CONFIDENCE. This results in a unique feature of each focal shop's customers' behaviors.

The pattern iv indicates that people who made transactions in each district and the focal store's customers behave in different manners, but such correlations may be by chance due to the lower CONFIDENCE. Pattern v indicates that people made transactions in each district and the customers of the focal store behave in a quite similar manner, meaning that the purchase behaviors of each focal shop's customers derive from the features of the shopping behaviors of customers who visited the district rather than the focal shop itself.

The next stage in analyzing the data was to attempt to discover the higher value of SUPPORT (>0.005) within patterns i and iii, which show both higher values of CONFIDENCE and the extremely high or low value of LIFT. SUPPORT generates the probability of a detected correlation over the total number of transactions made for an entire city. Thus, the score of SUPPORT indicates the strength of the pattern, considering the whole city rather than only a district.

The three categories with the red dots in the SUPPORT area in Figure 7 present the final identified unique patterns among stores. The categories include shop j's incoming customers in B1 and A2 and the shop k's outgoing customers in A2.

## 5. Discussion and conclusion

Urban association rules provide us with more insight into linked trips for shopping all over the city. Due to the introduction of the state-of-the-art-technologies in retail stores, the retail planning and marketing strategies inside stores are greatly improved (Pantano & Timmermans, 2011). However, the area development based on the analysis of linked trips for shopping is still limited. In addition, there is a lack of robust tools to enable us to analyze the large-scale datasets of people's purchase behaviors. Our proposed methodology can fill in these gaps.

The current study suggests the following direct benefits for transport and urban studies researchers:

(1) Five of six groups from three focal shops present similar behaviors with regard to the category B1 (hotel, hostel B&B and restaurant, pub and café), but the corresponding LIFT values vary greatly. For example, the analysis of LIFT indicates that there is a large difference in purchasing activities in this category between shop h's customers and people who live in or visit in the district where shop h is located. The former made many more transactions than the latter. This knowledge is useful, for example, because shop h's customers' origin and destination is known in terms of the shop's category, coupons or royalty cards can be introduced between the relevant stores to increase the number of transactions in this district. Conversely, shop h can provide similar coupons for their customers to use in stores of category B1 in other districts. Thus, city planners together with the commercial entities can attract or distribute more customers depending on the strategy of the district.

(2) Similarly, our proposed methodology is helpful for city authorities to evaluate the commercial planning for a specific district, overviewing its balance of an entire city. For example, in the district of Ciutat Vella, almost 38% of existing shops are classified into the category A2. However, the fraction between the number of stores of category A2 and the number of transactions conducted in this category is much less than 1.0 (i.e., 0.78). In addition, the ratio of the floor space of the existing store in this district and the number of transactions made in this district, again, is much lower than 1.0 (i.e., 0.55). Considering other factors such as the expenditure, this analysis is useful for city authorities to determine whether they should allow or limit the opening of new stores of the same or different categories within the area.

(3) More generally, the application of LIFT to a transaction dataset is helpful for creating an overview the situation of a city and the differences between districts. In the beginning of the planning process, the analysis of the districts that comprise a whole city is more important than the deep analysis of the districts in detail. That is, the quantity from making a larger number of proposals is more useful than the quality of each one in the early stages. Although causality must be explored, a quick draft should be made before performing a deep analysis. This greatly reduces the cost in terms of people and time for urban planners and city authorities. In this manner, this analysis enables us to evaluate several drafts and compare them for further more-elaborate planning.

(4) The dataset used for the proposed methodology derives from secondary data collected for other purposes, i.e., is a byproduct of people's activities. The transaction datasets are obtained when people make a transaction for the purchase of items in the store as a digital footprint or "data exhaust" (Mayer-Schönberger & Cukier, 2013, p113). This signifies that there is no need for further costs to collect the data for the linked trip analysis. The proposed method is highly helpful for transport and urban studies researchers or practitioners to continuously monitor and overview a person's cash flow in purchasing behaviors as a supplementary dataset without incurring additional costs.

Thus, the proposed method provides clear value and novel perspectives to the existing research, but it also has several limitations.

Our analysis is based on the successive order of customer purchases between different shops in which the customer uses BBVA's credit card. This means that we cannot detect transactions and interactions at locations where the customer does not use BBVA's credit card or does not do it through BBVA card terminals. In addition, our analysis does not consider the metric distance between stores, which is known as one of the important determining factors for shopping behaviors. Furthermore, our dataset cannot reveal the customer's decision-making process or value consciousness, which underlies the organization of activities in time and space, because it does not contain information about his/her inner thoughts, which is typically derived from interviews, questionnaires and participatory observation. Thus, the scope of this paper excludes generation of the daily activity-travel patterns of individuals, considering their socio-demographic characteristics. This indicates that we need to enrich the models via traditional data collection means such as household surveys. Moreover, our dataset contains a possible bias in terms of the representativeness of BBVA's customers relative to the economically active population in the given area. This potential bias is associated with the spatial inhomogeneity of BBVA market share. Thus, the obtained results in this paper are subject to further research or validation.

The application of association rules in the context of the city would allow researchers to create linked trips for shopping behaviors and thus create typologies of their activities. Although the methodology presented herein was applied to the particular selected stores, it could easily be generalized to other focal stores in an arbitrary urban context, and also to other equivalent types of data.

Overall, the method offers an effective means to extract the patterns of the linked trip for the shopping behaviors from bank card transactions and by doing so may also present new methods of analyzing such data. Our research considered the locations where customers made transactions rather than considering locations that the customers simply passed by. Most of the previous research, which used human movement datasets obtained using state-of-the-art-technologies, largely depended on inferring people's activities by combining other types of data (e.g., land use data) with the reconstructed paths (Alexander et al., 2015). Such datasets are not likely to contain people's expenditures during their trip. This is a piece of critical information that was not obtainable prior to this study.

# 8. Appendix

| Type | General Category | Sub-category | Activity | The number of transactions (%) | Classification of customers activities | | | | | | | |
|---|---|---|---|---|---|---|---|---|---|---|---|---|
| | | | | | Shop h | | Shop k | | Shop j | | | |
| | | | | | leaving | incoming | leaving | incoming | leaving | incoming | | |
| PRIM/SEC | A | - | Retail commerce | 62.6% | | | | | | | | |
| SEC | A | A1 | Convenience goods | 22.3% | v | iii | v | v | i | i | | |
| SEC | A | A2 | Comparison goods | 35.2% | v | v | i | v | v | v | | |
| SEC | A | A4 | Sell, fix and maintenance motor vehicles and fuel | 5.1% | v | v | v | iv | iii | iii | | |
| SEC | B | - | | 16.6% | | | | | | | | |
| SEC | B | B1 | Hotel, B&B, hostel, restaurant, pub, café | 16.6% | i | i | v | v | v | i | | |
| PRIM/SEC | C | - | IT tech, services to business and people, R&D | 17.0% | | | | | | | | |
| PRIM | C | C1 | Activities related to transport and travel | 11.6% | v | v | v | v | i | i | | |
| PRIM | C | C2 | Financial intermediation, except insurance | 1.4% | v | v | ii | v | v | v | | |
| PRIM | C | C3 | Insurance | 0.6% | v | v | v | v | v | v | | |
| PRIM | C | C4 | Activities related to financial intermediation | 0.4% | iv | iv | v | v | ii | v | | |
| PRIM | C | C5 | Real estate | 1.2% | iv | v | v | v | v | v | | |
| PRIM | C | C9 | Other service activities | 0.02% | iv | iv | iii | iv | ii | iv | | |
| SEC | C | C10 | Other services to people | 1.8% | iv | iv | v | iv | v | iv | | |

| | | | | Classification of customers activities | | | | | | |
|---|---|---|---|---|---|---|---|---|---|---|
| | | | | Shop h | | Shop k | | Shop j | | |
| Type | General Category | Sub-category | Activity | The number of transactions (%) | leaving | incoming | leaving | incoming | leaving | incoming |
| PRIM/SEC | E | - | PA, services of education, health and social assistance | 2.2% | v | v | v | v | v | v |
| SEC | E | E1 | Education | 0.5% | | v | v | v | v | v |
| PRIM | E | E2 | Health and social assistance | 1.7% | v | iv | ii | iv | iv | iv |
| PRIM | F | - | Associational, recreational and sport activities | 0.8% | | | | | | |
| PRIM | F | F1 | Recreational, cultural And sport activities | 0.8% | v | v | v | v | ii | v |